\documentclass[aps,prl,10pt,twocolumn,amsmath,amssymb,superscriptaddress]{revtex4-1}
\usepackage[]{graphicx}
\usepackage{comment}
\usepackage{color}
\usepackage{hyperref}
\usepackage{here}

\newcommand{\del}{\partial}

\newcommand{\tr}{\text{tr}}

\newcommand{\nn}{\nonumber}
\newcommand{\bbR}{{\mathbb R}}

\begin{document}

\begin{titlepage}

\title{Inequivalence between the Euclidean and Lorentzian versions of the type IIB matrix model from Lefschetz thimble calculations}

\author{Chien-Yu C\textsc{hou}}
\email{ccy@post.kek.jp}

\author{Jun N\textsc{ishimura}}
\email{jnishi@post.kek.jp}

\author{Ashutosh T\textsc{ripathi}}
\email{tripashu@post.kek.jp}

\affiliation{KEK Theory Center, Institute of Particle and Nuclear Studies,\
High Energy Accelerator Research Organization,\
1-1 Oho, Tsukuba, Ibaraki 305-0801, Japan}
\affiliation{Graduate Institute for Advanced Studies, SOKENDAI,\
1-1 Oho, Tsukuba, Ibaraki 305-0801, Japan}

\date{January 31, 2024; preprint: KEK-TH-2686}
%%\date{\today; preprint: KEK-TH-2686}

\begin{abstract}
  The type IIB matrix model
  is conjectured to describe superstring theory nonperturbatively in terms of ten $N \times N$ bosonic traceless Hermitian matrices $A_\mu$ ($\mu=0, \ldots , 9$), whose eigenvalues correspond to (9+1)-dimensional space-time. Quite often, this model has been investigated in its Euclidean version,
  %% obtained by making a ``Wick rotation'' $A_0=-i A_{10}$ with $A_{10}$ being Hermitian,
  which is well defined although the ${\rm SO}(9,1)$ Lorentz symmetry of the original model is
  replaced by the ${\rm SO}(10)$ rotational symmetry.
  Recently, a well-defined model respecting the Lorentz symmetry has been proposed
  by ``gauge-fixing'' the Lorentz symmetry nonperturbatively using the Faddeev-Popov procedure. 
  Here we investigate the two models by Monte Carlo simulations overcoming the severe sign problem by the Lefschetz thimble method, in the case of matrix size $N=2$ omitting fermionic contributions.
We add a quadratic term $\gamma \, \tr (A_\mu A^\mu)$ in the action
and calculate the expectation values of rotationally symmetric (or Lorentz symmetric)
observables as a function of the coefficient $\gamma$.
Our results exhibit striking differences between the two models around $\gamma=0$ and in the $\gamma>0$ region associated with the appearance of different saddle points, clearly demonstrating
their inequivalence
%% of the two models
against naive expectations from quantum field theory.
\end{abstract}

\maketitle
\end{titlepage}

\textit{Introduction---} 
The type IIB matrix model \cite{9612115}
(or the Ishibashi-Kawai-Kitazawa-Tsuchiya model)
has been attracting attention as a promising nonperturbative formulation of superstring theory analogous to the lattice gauge theory. In particular, not only space but also time is expected to emerge dynamically from the 10 bosonic $N \times N$ Hermitian matrices $A_\mu$ ($\mu = 0 , \ldots , 9$). From this viewpoint, the possibility of the emergence of $(3+1)$-dimensional space-time has been investigated
intensively.

Historically, the emergence of 4D space-time was first discussed in the Euclidean model, which is well defined \cite{9803117,0103159}
although the ${\rm SO}(9,1)$ Lorentz symmetry of the original model
is replaced by the ${\rm SO}(10)$ rotational symmetry.
%% and can be obtained from the Lorentzian model by deforming the integration contour as
%% (i) $A_0 = -i A_{10}$ and
%% %replacing
%% (ii) $A_\mu = e^{\frac{1}{8} \pi i} \tilde{A}_{\mu}$ ($\mu = 1 , \cdots , 10$),
%% where $\tilde{A}_\mu$ are Hermitian.
%% The latter (ii) is needed to make the integrand of the partition function
%% $e^{iS} \mapsto e^{-S_{\rm E}}$, where $S$ is the original action
%% with ${\rm SO}(9,1)$ Lorentz symmetry
%% and $S_{\rm E}$ is the corresponding Euclidean action
%% %% , which has
%% with ${\rm SO}(10)$ rotational symmetry.
%% %% $A_0 = e^{-\frac{3}{8} \pi i} \tilde{A}_{10}$ and $A_i = e^{\frac{1}{8} \pi i} \tilde{A}_{i}$ ($i = 1 , \cdots , 9$), where $\tilde{A}_\mu$ are Hermitian.
The configurations
%% of $\tilde{A}_\mu$
that give the minimum action are diagonal up to ${\rm SU}(N)$ symmetry, where the diagonal elements represent the 10D space-time coordinates. Based on the one-loop
perturbative calculation
%% effective action
around such configurations, it was argued that the fermionic zero modes may play a crucial role in collapsing
the emergent space-time
%%the eigenvalue distribution of the matrices
%% $\tilde{A}_\mu$
to lower dimensions \cite{Aoki:1998vn}.

The first nonperturbative calculation in the Euclidean model was carried out by the Gaussian expansion method, where it was shown that the ${\rm SO}(10)$ symmetry is spontaneously broken down to ${\rm SO}(3)$ \cite{1108_1293}. This conclusion was also confirmed
by numerical simulations \cite{2002_07410},
in which the phase of the Pfaffian that arises from integrating out fermionic matrices plays a crucial role \cite{0003223,Nishimura:2000wf}.
See Ref.~\cite{Anagnostopoulos:2022dak} for a review.
%%See Refs.~\cite{Anagnostopoulos:2022dak,Brahma:2022ikl,Klinkhamer:2022frp,Steinacker:2024unq} for related reviews and a textbook.

In the Lorentzian model,
%%on the other hand, 
non-diagonal saddle-point configurations that describe expanding space-time
were
%% found
shown to appear
if one adds a quadratic term (or the ``mass term'')
to the action \cite{Kim:2011ts,Kim:2012mw,Steinacker:2017vqw,Steinacker:2017bhb,Hatakeyama:2019jyw,Sperling:2019xar,Steinacker:2021yxt},
which is motivated as an infrared regularization \cite{1108_1540}.
%% , which is motivated as an infrared cutoff that regularizes the Lorentzian model.
Recent numerical studies \cite{Hirasawa:2024dht}
%% Hatakeyama:2022ybs,Nishimura:2022alt,Anagnostopoulos:2022dak}
%%
%%\cite{1904_05919,Hatakeyama:2022ybs,Nishimura:2022alt,Anagnostopoulos:2022dak}
also suggest the emergence of expanding space-time, 
where
%%and the possibility that
only three out of nine spatial directions expand if fermionic contributions
are included properly.
(See also Refs.~\cite{Brahma:2022ikl,Klinkhamer:2022frp} for recent work
related to cosmology.)
%% the possibility that (3+1)D expanding space-time emerges
%% in the Lorentzian model with the Lorentz invariant mass term.
%introducing a cutoff on the integration domain of $A_\mu$.

However, it was pointed out recently \cite{Asano:2024def}
that the partition function of the Lorentz symmetric model is \emph{necessarily} divergent
due to the noncompactness of the symmetry group,
%% If one regularizes this divergence by introducing a Lorentz symmetry breaking cutoff,
%% severe artifacts remain even if an appropriate limit is taken later.
%% The mass term introduced
%% in the previous simulations \cite{Hatakeyama:2022ybs,Nishimura:2022alt,Anagnostopoulos:2022dak}
%% for infrared regularization is actually Lorentz symmetric,
%% and it does not cure the divergence due to the noncompact symmetry.
%% [There, the Lorentz symmetry was actually practically broken
%% since it is impossible to sample all the Lorentz-boosted configurations with equal weights,
%% which implies the ergodicity problem.]
%% In view of this, it
and it was proposed to remove this divergence
%%the divergence due to the noncompactness of the symmetry group
by gauge-fixing the Lorentz symmetry
%% nonperturbatively
using the Faddeev-Popov procedure. 
Thus a totally well-defined model respecting the Lorentz symmetry was obtained.

In this Letter, we perform Monte Carlo simulations of this ``gauge-fixed''
Lorentzian model
for the first time in the case of matrix size $N=2$ omitting fermionic contributions.
Since the integrand of the partition function involves a phase factor,
%%$e^{iS}$,
there is a severe sign problem, which we overcome by the
Lefschetz thimble method (LTM)
%%\cite{alexandru2016signproblemmontecarlo,Fukuma:2019uot},
\cite{Witten:2010cx,Cristoforetti:2012su,Cristoforetti:2013wha,Fujii:2013sra,Alexandru:2015sua,Fukuma:2019uot}.
This method is based on firm mathematical grounds such
as Cauchy's theorem and the Picard-Lefschetz theory for deforming the integration
contour in the complex plane,
and it has been applied successfully to quantum cosmology \cite{Chou:2024sgk}.
%%(See Ref.~\cite{Chou:2024sgk} for an application to .)
%%, for instance.)
%% In particular,
%% unlike the more tricky complex Langevin method (CLM)
%% that has been used so far in simulating the type IIB related matrix models,
%% the LTM enables us to sum over contributions from all the relevant saddle
%% points and fluctuations around them.
Here we obtain explicit results for Lorentz invariant observables as a function
of the coefficient $\gamma$ of the Lorentz invariant mass term.
%%which enables us to identify the saddle points that dominate at large $|\gamma|$.

%%As a comparison, we
We also perform simulations of
the Euclidean model, which is obtained
from the gauge-unfixed Lorentzian model
by replacing $A_0 = -i A_{10}$.
%%Similarly to the case of the gauge-fixed Lorentzian model,
One might naively think that the two models are equivalent since they are
related to each other by Wick rotation, which is commonly used in quantum field theory. 
However, our results exhibit striking differences between the two models,
in particular,
around $\gamma=0$ and in the $\gamma>0$ region associated with
the appearance of different saddle points.

%%\textit{Saddles in the type IIB matrix model---} 
\textit{Saddles in the gauge-unfixed model---} 
The partition function of the type IIB matrix model can be formally
written as
\begin{align}
  Z = \int d A e^{iS_{\rm b}[A]} \, {\rm Pf}{\cal M}[A] \ ,
  \label{def-Z-ikkt}
\end{align}
where $A_\mu$ ($\mu = 0 , \ldots , 9$) are $N\times N$ traceless Hermitian matrices. The bosonic action $S_{\rm b}[A]$ is given by \cite{9612115}
\begin{align}
  S_{\rm b}[A] &= - \frac{1}{4} N  \tr [A_\mu , A_\nu][A^\mu , A^\nu] \ ,
  \label{bosonic-action}
\end{align}
where the indices $\mu$ and $\nu$
are raised and lowered using the Lorentzian metric $\eta_{\mu\nu}= {\rm diag} (-1 , 1 , \ldots , 1)$
and repeated indices are summed over.
The model has ${\rm SO}(9,1)$ Lorentz symmetry $A_\mu ' = {\cal O}_{\mu\nu} A_\nu$,
where ${\cal O}\in {\rm SO}(9,1)$,
as well as the ${\rm SU}(N)$ symmetry $A_\mu ' = U A_\mu U^\dag$, where $U\in {\rm SU}(N)$.
The Pfaffian ${\rm Pf}{\cal M}[A] \in \bbR$ in \eqref{def-Z-ikkt} represents the contribution from the fermionic matrices, which makes the model maximally supersymmetric.
In what follows, we omit the Pfaffian,
%%for simplicity
and generalize the model to $D=d+1$ dimensions, where $d$ represents the number of spatial matrices.

The classical equation of motion or the saddle-point equation is given by
\begin{align}
  [A_\nu , [A^\nu , A_\mu] ] &= 0
  \quad \quad \mbox{for all~}\mu=0, 1 , \ldots , d \ .
  \label{classical_eom}
\end{align}
One can easily prove that all the solutions are given by diagonal matrices up to the ${\rm SU}(N)$ symmetry.
(See Appendix A of Ref.~\cite{Steinacker:2017vqw}, for instance.)

If one adds the Lorentz invariant mass term
\begin{align}
  S_{\rm m}[A] &= - \frac{1}{2} \gamma N \, \tr (A_\mu  A^\mu)
%%   \ ,
  \label{Linv-mass-term}
\end{align}
to the action, the saddle-point equation becomes
\begin{align}
[A_\nu, [A^\nu, A_\mu]] = \gamma A_\mu 
\quad \quad \mbox{for all~}\mu=0, 1 , \ldots , d \ .
\label{classical_eom-gamma}
\end{align}
The solutions of this equation for $\gamma \neq 0$ can be written in general as $A_\mu = \sqrt{|\gamma|} A^{(0)}_\mu$, where $A^{(0)}_\mu$ is some configuration independent of $\gamma$.
Using the same rescaling, one also finds that the partition function \eqref{def-Z-ikkt} is dominated by some saddle-point configurations at large $|\gamma|$, where quantum corrections are suppressed by $1/\gamma^2$.
Thus the ``massless'' limit $\gamma \rightarrow 0$ corresponds to the strong coupling limit.

Solutions of Eq.~\eqref{classical_eom-gamma} were generated numerically \cite{Hatakeyama:2019jyw} and they typically represent expanding space-time for $\gamma>0$, where the number of expanding directions is left arbitrary at the classical level. For $\gamma < 0$, no such solutions were found.

\textit{Saddles in the gauge-fixed model---}
Recently \cite{Asano:2024def},
it was pointed out that the partition function \eqref{def-Z-ikkt} is actually divergent
since all the Lorentz-boosted configurations
%% related to each other by Lorentz transformation
contribute equally,
and that a naive cutoff that breaks the Lorentz symmetry leaves a severe artifact
even if one takes the limit of removing the cutoff eventually.
However, this divergence is unphysical in the sense that it is simply due to the
redundancy of description. Hence, a physically appropriate  way to get rid of this divergence is to
choose a representative of the configurations that are related to each other
by Lorentz transformation
and to integrate over the representative only.
This can be done by the Faddeev-Popov procedure,
which is used also in fixing the gauge in gauge theories.
The gauge-fixed model thus obtained is given by \cite{Asano:2024def} 
%%Asano:2024def}
\begin{align}
  Z &= \int d A \,
  {\Delta}_{\rm FP}[A]\,
  \prod_{i=1}^{d} \delta(\mathrm{tr}(A_0 A_i)) \,
 e^{i(S_{\rm b}[A]+S_{\rm m}[A])} \ ,
  \label{gauge-fixed-ikkt}
\end{align}
where the $\delta$-function represents
the gauge-fixing condition \(\mathrm{tr}(A_0 A_i) = 0\) for
\(i = 1 , \ldots , d \).
%%all \(i\).
${\Delta}_{\rm FP}[A]$ represents the corresponding 
Faddeev-Popov (FP) determinant
%% given by
\begin{align}
  {\Delta}_{\rm FP}[A] &= {\rm det} \,  \Omega \ , \quad
  \Omega_{ij} = \tr (A_0)^2 \delta_{ij} + \tr (A_i A_j) \ ,
  \label{def-FPdet}
\end{align}
where $\Omega$ is a $d \times d$ real symmetric matrix.

%%Let us consider the saddle points 
The saddle-point equation for the gauge-fixed model \eqref{gauge-fixed-ikkt}
is given by
%% \eqref{classical_eom-gamma}
%%becomes
\begin{align}
\label{classical_eom-fixed}
& N [A_\nu, [A^\nu, A_\mu]] = \gamma N A_\mu +
i {\rm Tr} \left( \Omega^{-1} \frac{\del \Omega}{\del A^{\mu}}\right)
\end{align}
for all $\mu=0, 1 , \ldots , d$.
Since the gauge-fixed model \eqref{gauge-fixed-ikkt} still has ${\rm SO}(d)$ rotational symmetry,
we use it to restrict ourselves to configurations with $\tr (A_i A_j)=0$ for $i \neq j$,
which reduces the saddle-point equation \eqref{classical_eom-fixed} to
\begin{align}
& N [A_\nu, [A^\nu, A_\mu]]
  = (\gamma N + i  \kappa_\mu) A_\mu \   \mbox{(no sum over $\mu$)} \ ,
  \label{classical_eom-gauge-fixed} \\
 & \kappa_i = 2 \,  \{ \tr (A_0)^2 + \tr (A_i)^2 \}^{-1} \ ,
\quad  \kappa_0 = \sum_{i=1}^d \kappa_i \ .
\label{def-kappa}
\end{align}
%%where no sum over $\mu$ is taken in \eqref{classical_eom-gauge-fixed}.
Thus we find that even for $\gamma=0$, we obtain a mass-like term
in \eqref{classical_eom-gauge-fixed} with the coefficients $\kappa_\mu$ determined
by \eqref{def-kappa}
in a self-consistent manner.
This implies, in particular, that saddle points
%%configurations
become non-diagonal even for $\gamma=0$.

Moreover, due to the ``$i$'' on the right-hand side of \eqref{classical_eom-gauge-fixed}, the saddle points become complex in general, which raises the issue of whether each saddle point is relevant in the sense of the Picard-Lefschetz theory. In this regard, we note that for large $|\gamma|$, the solutions of \eqref{classical_eom-gauge-fixed} reduce to the solutions of \eqref{classical_eom-gamma}.
%for the gauge-unfixed model.
Therefore, it is expected
that the relevant saddle points at large $|\gamma|$
in the gauge-fixed model
are given by complex solutions which are smoothly connected
to real solutions of \eqref{classical_eom-gamma}.
By ``real'', we actually mean \emph{Hermitian} $A_\mu$,
which has \emph{real} coefficients $A_\mu^a$ in the expansion
$A_\mu = \sum A_\mu^a t^a$ in terms of the ${\rm SU}(N)$ generators $t^a$.
The word ``complex'' is used similarly.
%%configurations represented by Hermitian matrices

\textit{The $N=2$ case---}
To be concrete, let us consider the $N=2$ case.
%%As a concrete example, let us consider the \(d=4\) and \(N=2\) case.
%%which can be generalized to higher-dimensional cases (\(d > 3\)).
In the gauge-unfixed Lorentzian model, the real solutions of
\eqref{classical_eom-gamma} for $\gamma>0$ are given by \cite{largeD}
\begin{eqnarray}
\mbox{(a)} \   A_\mu &=& 0  \ , \label{gauge-unfixed-solutions} \\
\mbox{(b)} \   A_\mu &=& \sqrt{\frac{\gamma}{4}}  \sigma_\mu \ (\mu=1 , 2)
    \ , \ A_\mu=0 \ \mbox{(otherwise)} \ , \nn \\
\mbox{(c)}  \   A_\mu &=& \sqrt{\frac{\gamma}{8}}  \sigma_\mu \ (\mu=1 , 2, 3) \ ,
  \ A_\mu=0 \ \mbox{(otherwise)} \ , \nn
  %& \quad A_3 = \cdots =  A_d = 0 \ ,
%  & \quad A_i =  0 \quad (i=3 , \ldots d) \ ,
\end{eqnarray}
where \(\sigma_i\) ($i=1,2,3$) are the Pauli matrices.
For $\gamma<0$, the trivial one (a) is the only real solution.
%% These solutions correspond to the relevant saddle points.

In the gauge-fixed Lorentzian model, on the other hand, 
the general solution of the saddle-point equations
\eqref{classical_eom-gauge-fixed} and \eqref{def-kappa}
can be parameterized as
%%\cite{Wang-private}
\begin{align}
&  A_0 = z \, \sigma_3 \ , \quad A_1 = x \, \sigma_1 \ ,  \quad  A_2 = y \, \sigma_2 \ ,
  \nonumber\\
&  A_i =  0 \quad (i=3 , \ldots , d) \ ,
  %& \quad A_3 = \cdots =  A_d = 0 \ ,
%  & \quad A_i =  0 \quad (i=3 , \ldots d) \ ,
\label{Pauli-general-solution}
\end{align}
%%up to the \({\rm SU}(2) \times {\rm SO}(d,1)\) symmetry,
up to the \({\rm SU}(2) \times {\rm SO}(d)\) symmetry,
where \(x, y, z \in \mathbb{C}\).
%% are complex variables,
Here we have assumed $d \ge 4$, in which case we need
%%In this case,
$A_0 \neq 0$ for finite $\gamma$
since otherwise $\Omega^{-1}$ in \eqref{classical_eom-fixed} becomes singular.
This plays an important role in the saddle-point structure
of the gauge-fixed Lorentzian model.
%%since $\Omega$ in \eqref{classical_eom-fixed} becomes singular
%%${\Delta}_{\rm FP}[A]=0$
%%This representation provides the general solution by explicitly reducing the matrix structure while preserving the system's essential dynamical features, constrained by the symmetries of the model. We leave the proof of generality in the full paper.

\begin{figure}[t]
    \centering
    \includegraphics[width=0.9\linewidth]{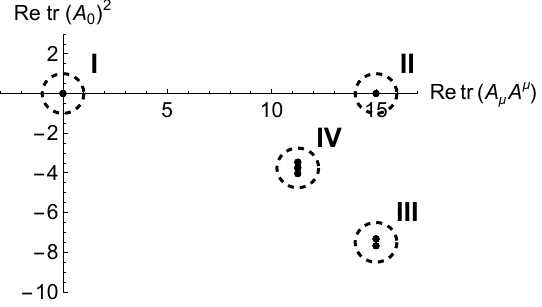}
    \caption{The nine saddle points for $\gamma=30$ are shown in the 
    $\mathrm{Re}\,\mathrm{tr}(A_\mu A^\mu)$--$\mathrm{Re}\,\mathrm{tr}(A_0)^2$ plane.
%%    The saddles are grouped according to their large $\gamma$ behavior. 
      There are actually three distinct points near the origin,
      which are close to each other.
    The dashed circles represent the group identification \eqref{saddle-groups-positive}.}
    \label{fig:saddle-class}
\end{figure}

Plugging \eqref{Pauli-general-solution} in 
\eqref{classical_eom-gauge-fixed} and \eqref{def-kappa}
and solving them for $x$, $y$, $z$,
we can obtain all the saddle points.
In Fig.~\ref{fig:saddle-class}, we show the nine saddle points for $\gamma=30$
in the $\mathrm{Re}\,\mathrm{tr}(A_\mu A^\mu)$--$\mathrm{Re}\,\mathrm{tr}(A_0)^2$ 
plane. 
The saddle points for $\gamma<0$ can be obtained from those for $\gamma>0$
with the same $|\gamma|$ by $A_\mu \mapsto i A_\mu$.
For instance, the plot for $\gamma=-30$ can be obtained
by rotating the one in Fig.~\ref{fig:saddle-class}
by 180 degrees.
%% According to the above discussions,
%% the candidates of relevant saddle points are given
%% by the solutions on the horizontal axis for $\gamma>0$, 
%% whereas for $\gamma<0$, they are given by the solutions near the origin only.

These nine saddle points
can be classified into the following four groups
according to
%%from
their asymptotic behaviors at large $|\gamma|$ up to symmetries.
%% There are actually nine of them, which
%% can be classified into the following four groups up to symmetries
%% from their asymptotic behaviors at large $|\gamma|$.
\begin{align}
      \mathrm{(I)} \ &  A_\mu \sim  0 \ ,  \nn \\
    \mathrm{(II)} \ &  A_\mu \sim \sqrt{\frac{\gamma}{4}}  \sigma_\mu \ (\mu=1 , 2)
    \ , \ A_\mu \sim 0 \ \mbox{(otherwise)} \ , \nn \\
    \mathrm{(III)} \ &  A_0 \sim  i \sqrt{\frac{\gamma}{4}}  \sigma_3 \ ,
    \ A_1 \sim \sqrt{\frac{\gamma}{4}}  \sigma_1 \ ,
    \ A_\mu \sim 0 \ \mbox{(otherwise)} \ , \nn \\
    \mathrm{(IV)} \ &  A_0 \sim i \sqrt{\frac{\gamma}{8}}  \sigma_3 \ , \
    A_\mu \sim  \sqrt{\frac{\gamma}{8}}  \sigma_\mu \ (\mu=1 , 2)\ , \nn \\
&     A_\mu \sim 0 \ \mbox{(otherwise)} \ .  
\label{saddle-groups-positive}
\end{align}
%% \begin{equation}
%%   \begin{array}{cll}
%%     \mathrm{(I)} & \quad \mathrm{tr}(A_\mu A^\mu) \sim 0 \ , &
%%          \mathrm{tr}(A_0^2) \sim  0 \ , \\
%%     \mathrm{(II)} & \quad \mathrm{tr}(A_\mu A^\mu) 
%%         \sim \tfrac{3}{4}\gamma \ , &
%% %% wrong!       \mathrm{tr}(A_0^2) \sim -\tfrac{1}{2}\gamma \ , \\
%%         \mathrm{tr}(A_0^2) \sim -\tfrac{1}{4}\gamma \ , \\
%%     \mathrm{(III)} & \quad \mathrm{tr}(A_\mu A^\mu) 
%%         \sim \gamma \ , &
%%         \mathrm{tr}(A_0^2) \sim -\tfrac{1}{2}\gamma \ , \\
%%     \mathrm{(IV)} & \quad \mathrm{tr}(A_\mu A^\mu) 
%%         \sim \gamma \ , &
%%         \mathrm{tr}(A_0^2) \sim 0 \ .
%%         \end{array}
%% \label{saddle-groups-positive}
%% \end{equation}
%%%

%% \begin{align}
%%     \mathrm{(I)} & \quad \mathrm{tr}(A_\mu A^\mu) \sim 0 \ , \quad
%%          \mathrm{tr}(A_0^2) \sim  0 \ , \nonumber\\
%%     \mathrm{(II)} & \quad \mathrm{tr}(A_\mu A^\mu) 
%%         \sim \tfrac{3\gamma}{4} \ , \quad 
%%         \mathrm{tr}(A_0^2) \sim -\tfrac{\gamma}{2} \ , \nonumber\\
%%     \mathrm{(III)} & \quad \mathrm{tr}(A_\mu A^\mu) 
%%         \sim \gamma \ , \quad 
%%         \mathrm{tr}(A_0^2) \sim -\tfrac{\gamma}{2} \ , \nonumber\\
%%     \mathrm{(IV)} & \quad \mathrm{tr}(A_\mu A^\mu) 
%%         \sim \gamma \ , \quad 
%%         \mathrm{tr}(A_0^2) \sim 0 \ .
%% \label{saddle-groups-positive}
%% \end{align}

For $\gamma>0$,
the solutions that become real at large $\gamma$
are those in groups (I) and (II),
which are smoothly connected to the solutions
(a) and (b), respectively, in the gauge-unfixed model.
%% at large $\gamma$.
There is no solution that is smoothly connected to (c)
%%, however,
as anticipated from \eqref{Pauli-general-solution}.
For $\gamma<0$, the solutions that become real at large $|\gamma|$
are those in group (I),
which are smoothly connected to the solution
(a) in the gauge-unfixed model.
%% at large $|\gamma|$.
These solutions are
%%the candidates for
the relevant saddle points at large $|\gamma|$.
By changing $\gamma$ continuously,
we find that relevant saddles at $\gamma>0$ are smoothly connected
to irrelevant saddles at $\gamma<0$ and vice versa, meaning that the group
identification of each saddle can change as we cross $\gamma=0$.

\textit{Defining the Euclidean model---}
So far, we have discussed the gauge-fixed Lorentzian model \eqref{gauge-fixed-ikkt},
which uses gauge fixing
to remove the divergence of the original model due to the noncompact symmetry group.

An alternative way to render
the partition function finite
%%make the type IIB matrix model well defined
is to
make a ``Wick rotation'' $A_0=-i A_{D}$ with $A_{D}$ being Hermitian.
The Euclidean model obtained in this way is
given by
\begin{align}
  Z_{\rm E} &= \int d A \, e^{i(S_{\rm b}[A]+S_{\rm m}[A])} \ ,
  \label{Euclidean-ikkt}
\end{align}
where the two terms $S_{\rm b}[A]$ and $S_{\rm m}[A]$ in the action
are formally the same as \eqref{bosonic-action} and \eqref{Linv-mass-term},
assuming that $A$ 
%%in \eqref{Euclidean-ikkt}
represents $A_\mu$ ($\mu = 1, \ldots , D$) and defining $A^{D}\equiv A_D$.
Since the ${\rm SO}(d,1)$ Lorentz symmetry of the original model is converted into
the ${\rm SO}(D)$ rotational symmetry,
there is no more divergence associated with the noncompactness of the symmetry group.
In the $N=2$ case, the saddle points are essentially the same as in the
gauge-unfixed Lorentzian model given by \eqref{gauge-unfixed-solutions} and below.

Note that the partition function of this Euclidean model still involves a
phase factor in the integrand.
The Euclidean model discussed in the literature involves $e^{-S_{\rm b}}$ instead,
and it can be obtained
by setting $\gamma=0$ and making a contour deformation
$A_\mu = e^{\frac{1}{8} \pi i} \tilde{A}_{\mu}$ ($\mu = 1 , \ldots , D$)
with $\tilde{A}_{\mu}$ being Hermitian.
The model \eqref{Euclidean-ikkt} may be viewed as a natural extension
of this conventional Euclidean model to $\gamma \neq 0$, which may be
compared directly with the gauge-fixed Lorentzian model
\eqref{gauge-fixed-ikkt} with the same $\gamma$.
%%for its obvious relationship.

\textit{Monte Carlo simulations---}
Let us discuss our results
%%of Monte Carlo simulations
based on the LTM
for the gauge-fixed Lorentzian model and the Euclidean model
%%gauge-unfixed (\( SO(d+1) \))
in the $N=2$ and $d=4$ case.
%% , omitting fermionic contributions.
%% models to clarify the roles of the different saddle points and examine how the parameter \(\gamma\) influences the dynamics.

\begin{figure}[t]
    \centering
    \includegraphics[width=0.9\linewidth]{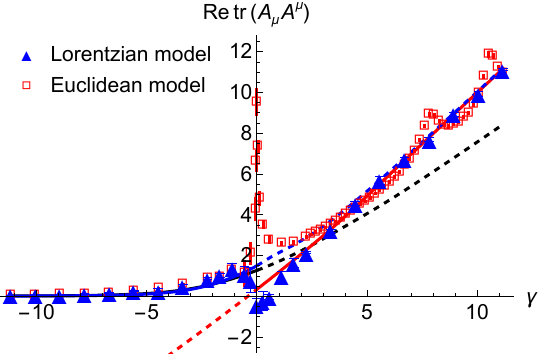}\\
    \includegraphics[width=0.9\linewidth]{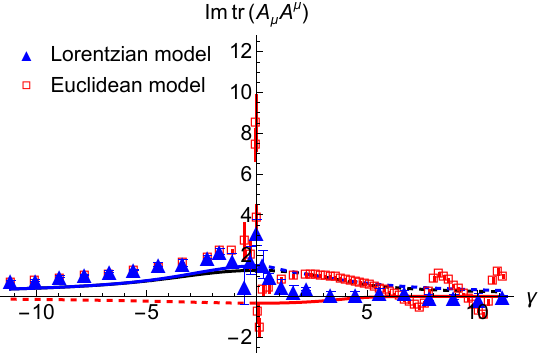}
    \caption{The real part (Top) and the imaginary part (Bottom) of
      \( \langle \mathrm{tr}(A_\mu A^\mu) \rangle \) are plotted against $\gamma$
      for the gauge-fixed Lorentzian model (blue triangles)
      and the Euclidean model (red squares).
      The red line represents the result obtained for
      the relevant saddle in group (II) at $\gamma>0$,
      while the black line and the blue line
      represent the results obtained for the relevant saddles in group (I) at $\gamma<0$.
      Each of these relevant saddles
      continue to irrelevant saddles as one crosses $\gamma=0$,
      which are indicated by the dashed line with the same color.}
    \label{fig:ReAndImA2}
\end{figure}

Fig.~\ref{fig:ReAndImA2} shows \( \langle \mathrm{tr}(A_\mu A^\mu) \rangle \)
as a function of \(\gamma\).
%Let us first discuss the behaviors in the \(\gamma > 0\) region.
First we discuss the results for the gauge-fixed Lorentzian model \eqref{gauge-fixed-ikkt}.
At large $|\gamma|$, the results are expected to approach the results for
some relevant saddle points discussed earlier.
For $\gamma>0$, the data points approach the red line 
representing the result for the saddle point in group (II) in \eqref{saddle-groups-positive}.
This saddle point continues to the one in group (IV) in the $\gamma<0$ region,
which is irrelevant.
For $\gamma<0$, the data points approach the blue line and the black line
representing the results for the two saddle points
in group (I) in \eqref{saddle-groups-positive}.
[The result for the third saddle point in group (I), which is not shown here,
diverges at $\gamma=0$.]
%%and we consider it to be irrelevant.]
The blue and black lines continue to the saddle points in groups (III) and (IV)
in the $\gamma>0$ region, respectively, which are irrelevant there.
Thus we observe a switching of the (dominant) relevant saddle points as we cross $\gamma=0$.
At $\gamma \sim 0$, we observe some oscillating behavior, which may be due to interference
between different saddles.

%% Taking this into account, one obtains the correct behavior \(-\frac{15i}{2\gamma}\),
%% which is different from the behavior \(-\frac{4i}{\gamma}\) obtained naively
%% from the saddle point. This is demonstrated 

Let us turn our attention to the results for
the Euclidean model \eqref{Euclidean-ikkt}.
In the $\gamma>0$ region, we see a clear oscillating behavior,
which is due to the interference between the two dominant saddle points
(b) and (c) in \eqref{gauge-unfixed-solutions}.
The partition functions $Z_{\rm b}$ and $Z_{\rm c}$
around these two saddles can be calculated perturbatively \cite{largeD},
and their ratio is given by \(  Z_{\rm b}/ Z_{\rm c} \sim 2 e^{-i\gamma^2/8}\) at large \(\gamma\).
We can therefore make an estimate
\begin{align}
  \langle \mathrm{tr}(A_\mu A^\mu) \rangle \sim
  \frac{  \langle \mathrm{tr}(A_\mu A^\mu) \rangle_{\rm b} Z_{\rm b}
+ \langle \mathrm{tr}(A_\mu A^\mu) \rangle_{\rm c} Z_{\rm c}}{Z_{\rm b} + Z_{\rm c}} \ , 
  \end{align}
with $\langle \mathrm{tr}(A_\mu A^\mu) \rangle_{\rm b}  \sim \gamma $ and
$\langle \mathrm{tr}(A_\mu A^\mu) \rangle_{\rm c}  \sim \frac{3}{4}\gamma $,
which roughly explains the oscillating behavior.

\begin{figure}[t]
    \centering
    \includegraphics[width=0.9\linewidth]{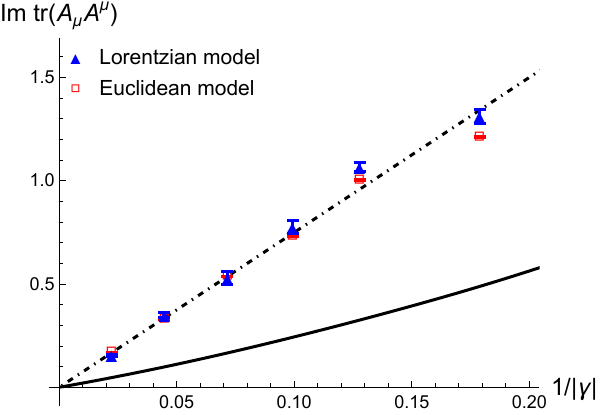}
    \caption{The imaginary part of
          \( \langle \mathrm{tr}(A_\mu A^\mu) \rangle \)  
%%    \(\mathrm{Im}\,\mathrm{tr}(A_\mu A^\mu)\) versus \(1/\gamma\) 
          is plotted against \(1/|\gamma| \)
          for the gauge-fixed Lorentzian model (blue triangles) and
          the Euclidean model (red squares).
          The dash-dotted line represents 
    the leading behavior \( \frac{15/2}{|\gamma|} \) 
%%    \(\langle\mathrm{Im}\,\mathrm{tr}(A_\mu A^\mu)\rangle = -15/(2\gamma)\) 
    obtained by omitting the quartic term $S_{\rm b}$.
          The solid line represents the leading behavior \( \frac{4}{|\gamma|}\)
          obtained from the saddle points in group (I).
          %%in the gauge-fixed Lorentzian model.
    }
    \label{fig:inversegamma}
\end{figure}

In the $\gamma<0$ region,
the data points
%%or the Euclidean model and the gauge-fixed Lorentzian model
for the two models
get close to
each other as $|\gamma|$ increases.
This is understandable since the dominant saddle points are
essentially the trivial one $A_\mu = 0$
in the $\gamma \rightarrow - \infty$ limit for both models.
%%the $\gamma$ dependence of
In fact, the asymptotic behavior of the partition functions
\eqref{gauge-fixed-ikkt}
and \eqref{Euclidean-ikkt}
%%The asymptotic behavior
%%for $\gamma \rightarrow - \infty$
is readily obtained by just omitting the quartic term $S_{\rm b}$
and rescaling $A_\mu \mapsto A_\mu/\sqrt{|\gamma|}$,
which yields the common result $Z \sim |\gamma| ^{-3D/2}$.
One can then obtain $\langle \mathrm{tr}(A_\mu A^\mu) \rangle
= i  \frac{d}{d\gamma} \log Z = \frac{(3D/2)i }{|\gamma|}$.
%for both the gauge-fixed Lorentzian model and the Euclidean model in $D$ dimensions.
In Fig.~\ref{fig:inversegamma},
we plot the imaginary part of \( \langle \mathrm{tr}(A_\mu A^\mu) \rangle \)
against $1/|\gamma|$ for the two models, which shows
the same asymptotic behavior as predicted.
%% agrees with the predicted asymptotic behavior.
On the other hand, the two models
start to show different behaviors at $|\gamma| \lesssim 10$,
which suggests that the two models agree only in
the leading asymptotic behavior at $\gamma \rightarrow - \infty$.
%%
%% discrepancies
%% showing up already at $|\gamma| \lesssim 10$,
%% which suggest that the agreement holds only in the leading behavior.
%%at the leading order in the $1/\gamma^2$ expansion.

Incidentally, 
let us note that the asymptotic behavior obtained here
cannot be reproduced by the relevant saddle points
in the gauge-fixed Lorentzian model in the $\gamma<0$ region. (See Fig.~\ref{fig:inversegamma}.)
%%which belongs to group (I) in \eqref{saddle-groups-positive}.
This is due to the fact that the Faddeev-Popov determinant ${\Delta}_{\rm FP}[A]$
in \eqref{gauge-fixed-ikkt} becomes zero for the saddle points in group (I)
at $\gamma\rightarrow - \infty$.

%% \begin{figure}[b]
%%     \centering
%%     \includegraphics[width=0.8\linewidth]{img/commute.pdf}
%%     \caption{The quantity $\rho$ defined in \eqref{def-rho}
%% %%    \(\left|\frac{-\,\mathrm{tr}([A_\mu, A_\nu][A^\mu, A^\nu])}{
%%       %%     (\mathrm{tr}(A^\mu A_\mu))^2}\right|\) versus \(\gamma\)
%%       is plotted against $\gamma$ for the Euclidean model.}
%%     %%   This quantity drops sharply near \(\gamma=0\),
%%     %%   reflecting an emergent commuting saddle 
%%     %% at \(\gamma=0\).}
%%     \label{fig:Commute}
%% \end{figure}

Finally let us comment on the prominent peak observed in Fig.~\ref{fig:ReAndImA2}
around $\gamma = 0$ for the Euclidean model.
%%We infer that
This is actually due to the emergence of commuting saddles
represented by diagonal solutions to \eqref{classical_eom} at $\gamma=0$,
which do not occur
%%have any counterparts
in the gauge-fixed Lorentzian model.
This can be confirmed by calculating a quantity
\begin{align}
\rho =   \left|  \frac{\left\langle-\,\mathrm{tr}([A_\mu, A_\nu][A^\mu, A^\nu]) \right\rangle }{
  \left\langle\mathrm{tr}(A_\mu A^\mu)\right\rangle ^2}  \right|
%% \ ,
\label{def-rho}
\end{align}
that probes
the noncommutativity of the dominant configurations,
which shows a sharp dip around $\gamma =0$ as a function of
$\gamma$ \emph{only} in the Euclidean model \cite{full-paper}.

%% In order to confirm this numerically, we calculate the quantity
%% \begin{align}
%% \rho =   \left|  \frac{\left\langle-\,\mathrm{tr}([A_\mu, A_\nu][A^\mu, A^\nu]) \right\rangle }{
%%   \left\langle\mathrm{tr}(A_\mu A^\mu)\right\rangle ^2}  \right|  \ ,
%% \label{def-rho}
%% \end{align}
%% which probes the noncommutativity of the dominant configurations,
%% where the denominator is intended to make the quantity scale
%% invariant (\emph{i.e.}, under $A_\mu \mapsto c A_\mu$).
%% In Fig.~\ref{fig:Commute}, we plot the quantity $\rho$
%% against $\gamma$ for the Euclidean model, which clear shows a sharp dip
%% around $\gamma =0$.
%% The nonzero value of $\rho$ at $\gamma=0$ agrees with
%% the previous result for the conventional Euclidean model \cite{9811220},
%% and it should be attributed to quantum corrections.

\textit{Discussions---}
In quantum field theory, the Euclidean theory obtained by the Wick rotation
%% is often used to investigate the properties of the vacuum and
%% the excitations around it
%% through the calculations of
%% vacuum expectation values and correlation functions.
%%
%% This can be extended to the studies of
%% the equilibrium state at finite temperature.
%%
%% The Euclidean theory is also used in investigating quantum tunneling semi-classically
%% through instantons, and such calculations are extended even to quantum gravity.
%%
%%The Euclidean theory also
forms the basis of
the lattice gauge theory,
which plays an important role in defining gauge theories nonperturbatively.
%%the nonperturbative definition of gauge theories.
%%
%% For these reasons, we tend to consider that the Euclidean theory and the Lorentzian theory
%% are equivalent.
This point of view has been taken in the type IIB matrix model,
and the Euclidean version has been studied
intensively in the literature \cite{9803117,0103159,Aoki:1998vn,1108_1293,2002_07410,0003223,Nishimura:2000wf,Anagnostopoulos:2022dak}.
%%in most cases.

In this Letter, we have investigated the impact of the Wick rotation
on the nonperturbative dynamics of the type IIB matrix model.
In particular, this problem has become clearer than ever,
now that we have two well-defined models, namely
the Euclidean model with the ${\rm SO}(D)$ rotational symmetry on one hand
and the Lorentzian model with the ${\rm SO}(d,1)$ Lorentz symmetry
on the other hand \cite{Asano:2024def}.
These two models are related to each other by the Wick rotation $A_0=-i A_{D}$,
which may be viewed as a contour deformation $A_0=e^{-i \theta}  A_{D}$
from $\theta=0$ to $\theta=\frac{\pi}{2}$.
The interpolating model with $ 0 \le \theta < \frac{\pi}{2}$ is well defined
for $\gamma \le 0$ but not for $\gamma>0$.
Also there is a subtlety in the $\theta \rightarrow \frac{\pi}{2}$ limit
due to the emergence of Lorentz symmetry, which is noncompact \cite{Asano:2024def}.
%% In defining the latter, the Lorentz symmetry has to be gauge-fixed properly
%% by the Faddeev-Popov procedure in order to remove the apparent divergence due to
%% the noncompact symmetry group \cite{Asano:2024def}.
Whether the model defined by gauge-fixing the Lorentz symmetry
at $\theta=\frac{\pi}{2}$ is equivalent
to the Euclidean model at $\theta=0$ is therefore a nontrivial issue.

In order to give a clear answer to this equivalence issue, we have investigated
the Euclidean and Lorentzian models in the simplified case.
%% adding a mass term, which seems to play an important role
%% in the emergence of space-time.
%% In the simplest $N=2$ case omitting the fermionic contributions,
%% we have obtained nonperturbative results as a function of $\gamma$
%% %%investigate the models
%% by Monte Carlo simulations
%% overcoming the severe sign problem by the LTM.
While the two models show the same asymptotic behavior for $\gamma \rightarrow -\infty$
due to the dominance of the trivial saddle $A_\mu=0$ there,
otherwise they give totally different behaviors in the whole region of $\gamma$.
In particular, in the $\gamma > 0$ region, the saddle points (b) and (c) in
\eqref{gauge-unfixed-solutions} dominate in the Euclidean model,
whereas only the saddle point (b) dominates in the Lorentzian model.
At $\gamma\sim 0$, the commuting saddles appear in the Euclidean model,
whereas they do not show up in the Lorentzian model.
Thus our results clearly demonstrate that the two models are inequivalent.
%% which is not only surprising from what we know in quantum field theory
%% but also of great importance in the discussion of the emergent space-time
%% in the type IIB matrix model.

While the inequivalence between the Euclidean and Lorentzian models
has been demonstrated in the $N=2$, $d=4$ case omitting the fermionic
contributions for simplicity,
it should also be true in the type IIB matrix model at large $N$, 
which is relevant to nonperturbative formulation of superstring theory.
Also the emergence of nontrivial saddle points in the simple model
may be viewed as a prototype of the emergent space-time
in the type IIB matrix model.
We are currently extending our simulations to
larger $N$ and to the supersymmetric case using LTM, where 
the computational cost grows only by some power of $N$
instead of the exponential growth anticipated naively for a system with the sign problem.

Recently the Euclidean model with supersymmetric mass deformation
\cite{Bonelli:2002mb}
has been attracting attention
(See Refs.~\cite{Kumar:2022giw,Kumar:2023nya} for preliminary numerical results.)
since its gravity dual has been
identified \cite{Hartnoll:2024csr,Komatsu:2024bop,Komatsu:2024ydh}.
This gives a new perspective to the emergence of space-time in the type IIB
matrix model. It would be also interesting to investigate this model and its
Lorentzian version using the LTM.
%%by numerical simulations.
%%Extending our numerical simulations to larger $N$ and to the SUSY case
Another important direction is to identify the gravity dual of the Lorentzian models.
%%Identifying the gravity dual of the Lorentzian models is another challenge.

\textit{Acknowledgments---}
%\paragraph*{Acknowledgments.---}
We thank Yuhma Asano,
%Hikaru Kawai,
Worapat Piensuk
%, Harold Steinacker
and Naoyuki Yamamori for valuable discussions on the related
project \cite{largeD},
and Cheng-Tsung Wang for helpful discussions
%% his proof \cite{Wang-private}
on the form \eqref{Pauli-general-solution} of general solutions. 
This work was supported by JST, the establishment of university fellowships towards the creation of science
technology innovation, Grant Number JPMJFS2136.

\bibliography{thimble-ikkt}

%% \begin{thebibliography}{99}
%% \bibitem{Hartle:1983ai}
%% J.B.~Hartle, S.W.~Hawking, {\it {Wave Function of the Universe}},  {\em Phys. Rev. D} {\bf
%%   28} (1983) 2960--2975.
%% \end{thebibliography}

\end{document}